\documentclass[a4paper,oneside,12pt]{amsart} %
\usepackage{amsfonts}
\usepackage{amssymb}
\usepackage{amsmath}
\usepackage{amscd}
\usepackage{dsfont}
\usepackage{cite}
\usepackage{euscript}
\usepackage[dvips]{graphicx}
\usepackage{color}
\usepackage[table,xcdraw]{xcolor}
\usepackage[foot]{amsaddr}

\usepackage{cmbright}
\usepackage{mathrsfs}
\usepackage[utf8]{inputenc}
\usepackage[T1]{fontenc}

\numberwithin{equation}{section}

\newcommand{\abs}[1]{\left\vert#1\right\vert}

\newcommand{\eps}{\varepsilon}

\renewcommand{\i}{\mathrm{i}}
\newcommand{\tr}{\textrm{Tr}\;}

\renewcommand{\Im}[1]{\mathbb{I}\mathrm{m}\left\{#1\right\}}

\newcommand{\bm}{\boldsymbol}

\newcommand{\bmD}{\boldsymbol D}

\newcommand{\bmP}{\boldsymbol P}

\title[The minimal model of LH2 from ab initio calculations]{The minimal model of light harvesting complex with dipole-quadrupole interaction derived from ab initio calculation.}
\author[1]{ V. Al. Osipov$^*$}
 \author[2]{Luca De Vico$^{**}$}
\author[3]{Andr\'e Anda$^\dag$}
\author[4]{Thorsten Hansen$^\dag$}
 \address[*]{Department of Chemical Physics, Lund University, Getingev\"agen 60, 22240 Lund, Sweden}
 \address[$**$]{Department of Biotechnology, Chemistry and Pharmacy,
University of Siena, via A. Moro 2, I-53100 Siena, Italy}
\address[$\dag$]{Department of Chemistry, University of Copenhagen, Universitetsparken 5, DK-2100,
Copenhagen $\emptyset$, Denmark}

\date {\today}

\begin{document}
\maketitle
\begin{abstract}
A minimal, one-parameter model of the excitonic Hamiltonian for the light harvesting complex of purple bacteria (LH2) based on ab initio calculation of the excitonic energies is proposed. The set of input parameters contains positions of atoms in 27 bacteriochlorophyll pigments only. The excitation energies, transition dipole and quadrupole moments of the bacteriochlorophyll units were calculated by advanced multiconfigurational multireference method. In the work we investigate influence of the dipole-quadrupole interaction on the particularities of the absorption spectra and circular dichroism and on the wave-functions localization. We demonstrate that although the spectra changes are small the dipole-quadrupole interaction term leads, however, to almost complete separation of the system onto two subsystems: the small B800 ring  of pigments and the large B850 ring. The latter observation is seemed essential for formulation of energy transport models in light harvesting complexes.
\end{abstract}

\setlength{\textfloatsep}{20pt plus 4pt minus 4pt}
\setlength{\floatsep}{20pt plus 4pt minus 4pt}
\setlength{\columnsep}{9pt}

\section{Introduction} The basic energy source for plants is photosynthesis. During the last few decades understating of the energy transfer mechanism in all details was a subject of large scientific interest inspired by the fundamental role of the process in biology. Photosynthetic systems have developed various antenna systems. The light-harvesting (LH) antenna capture sunlight with very high efficiency and successfully transfers the energy of light-excitation to the photosynthetic reaction center~\cite{PS1996, DPZ2016}, where it activates the electron-transfer reaction to produce ATP, the molecule serving for purposes of intracellular transport of chemical energy required for metabolism. The molecular architecture of the antenna has been explored by the methods of X-ray diffraction in great details~\cite{RHSGLIC2003, PPHCI2003, NYTHKWM2014}. This information have opened a possibility to describe the system on the atomic level and, being supplemented by spectroscopic data~\cite{KZFG1997, WRS1997, KTSCKF2013, GFBSRC1999}, eventually resulted in a number of theoretical concepts addressing the electronic properties of the system. The importance of such efforts and their ultimate goal is evident. It is  not only {\it ab initio} calculation of the spectral features, but also access to the fundamental relationship between the protein structure and its biological functions. 

In this article our attention is focused on the study of a classical object of investigation in photosynthesis, the LH complex of the purple bacteria ({\it Rhodoblastus acidophilus}). The LH complex contains two major types of antenna proteins. One of these (LH1) is known to surround the reaction center, while the other smaller proteins (LH2) form a peripheral network. The structure of the trans-membrane protein LH2 is highly symmetric. It consists of an outer and inner cylinders made up of $\alpha$-helices, which together hold two rings of bacteriochlorophyll (BChl) units and carotinoids. One of the rings contains 9 units (along the text they will be marked by letter $\gamma$) positioned almost perpendicularly to the cylinders' guides. The other ring consists of 9 pairs of alternatingly arranged units, in each pair one unit is anchor to the inner ($\alpha$-unit) and the other to the outer ($\beta$-unit) protein cylinder. A circle of eight carotenoids bridges the rings, such that each carotinoide is in contact with one $\gamma$ and one $\alpha$ unit. Both rings are located in the apolar part of the protein, so that the electrostatic effect from the protein environment are assumed to be negligible. The essential influence, however, comes from even small conformational rearrangements, which give uncontrolled contribution to the interactions between the BChl units and thus introduce both static and dynamic disorder into the system. The reorganisation of atomic structure after excitation of the electron density enters as a perturbation to the diagonal elements of the excitation Hamiltonian. 

The absorption spectra of LH2 complex differ substantially from the one of monomeric BChl's. The long-wavelength ($Q_y$) absorption band of BChl occurs at around 780~nm and has a negligible rotational strength. The excitation process results in change of the electronic density distribution, no bond dissociation occurs and the nuclear positions in the excited states are only slightly different from those in the ground state~\cite{S1975}. The redistribution of density mostly confined to $\pi$-electrons from the conjugated porphyrin ring system. The BChl units gathered into the LH2 complex give rise to two absorption bands. One occurs at around 800 nm.  This band is associated with the $\gamma$ ring of BChl units~\cite{GFBSRC1999, GFJKCGZ2002,KTSCKF2013}, which is usually denoted as B800. The closer packing in the second ring shifts the absorption band to around 850 nm at room temperature and up to 870 nm at 4~K~\cite{WRS1997, GFJKCGZ2002,KTSCKF2013}. The 850 absorption peak is generally about twice that of the 800 band. The circular dichroism (CD) spectra display a negative lob on the red side of the 850-nm band and positive on the blue side and a somewhat weaker pair of lobes with reversed sign on the 800-nm band~\cite{GFJKCGZ2002}. The CD spectra show more variations with respect to the experimental conditions. This statement has been also supported by by the numerical study, and is shown to be sensitive to the relative orientations of the transition dipole moments of BChl units~\cite{CJCCGKGCM2016}. The most notable feature is the red shift of the zero crossing of the CD spectrum at 850 band with respect to the absorption maximum of around 6 nm, which was shown to be an indication of the energy difference between the $\alpha$ and $\beta$ units~\cite{KZFG1997}.

In description of the LH system one can follow two different strategies. The one is based on  adaptation of a suitable phenomenological model in order to reproduce the experimentally observed spectral features. The apologists of the other strategy starts from crystallographic data and try to predict the observables using general principles of quantum and statistical theory in their calculation program. Within the phenomenological picture aiming to describe the energy migration in LH2, the energy transfer is understood as F\"orster incoherent hopping between the weakly coupled units of B800 ring and between the rings, while transport in the dense B850 is considered as a migration of a molecular exciton~\cite{TMCM2000}. Thus, the central element for the quantitative description, is calculation of the coupling constants. From the particularities of the spectrum, namely separation of the bands from the rings B800 and B850, one can deduce that the largest constants are those that correspond to the dimeric coupling of the nearest $\alpha$ and $\beta$ units, with the mutual distance of around 9 \AA. Calculated in such semi-phenomenological way the values of these constants varies~\cite{TMCM2000, KSF1998, ZNC2004, CZHS1998} in the range from 250 to 800 cm$^{-1}$. 

The LH2 system have been also extensively studied by means of numerical methods~\cite{GFJKCGZ2002, NG2013, Luca}. Note, however, that application of the standard tools of quantum chemistry, such as density functional theory (DFT), for the calculation of coupling constants appeared to be problematic as well, since all they required an external input. For instance, it is not possible to calibrate the time-dependant-DFT method in terms of functionals and basis sets to obtain the desired couplings, out of lack of experimental data for the BChl embedded in LH2. In this article we use the results obtained by advanced multiconfigurational multireference (MCMR) method recently developed in the work~\cite{Luca}, which computational part is based completely on the crystallographic data and basic principles of quantum chemistry. Thus the method is free of any fitting parameters, moreover up to now this is the most exact calculation of the excitation spectra of BChl unit embedded into the protein complex. The method is utilised to compute the excitation energies of a single unit from ab initio modelling. However, calculation of the whole antenna complex in the same manner is still an unaffordable task and one has to follow the old strategy. The latter assumes formulation of a modelling one-exciton Hamiltonian with a field-matter interaction term and calculation out of it the spectral observables, such as susceptibility and circular dichroism, according to the known formula.

In this work we construct an ab initio model and formulate a minimal model based on the data obtained from MCMR for calculation of absorption spectrum and circular dichroism spectra. In addition to the known approaches, which include only the dipole-dipole interaction into the molecular Hamiltonian, we consider also the dipole-quadrupole correction to the interaction constants and discuss its influence on the spectral particularities. Note, that the need in accounting of the higher moments at description of electronic excitation interactions between large molecules has been noticed long time ago~\cite{C1977}, while never been included into models of LH2 complexes. Moreover, in the recent research~\cite{CJCCGKGCM2016}, discussing influence of temperature fluctuations on the spectrum, among others, a large sensitivity to even tiny changes in the relative orientation of the transition dipole moments of the interacting BChl units has been found. This forced us to think that the dipole-quadrupole contribution can give a comparable effect on the spectra. 

The article is organised in the following way: the next section is devoted to the mathematical concept used in our analysis; The results of our calculations are discussed in details in the last section. The article is supplemented by two appendices, where we have gathered the data obtained by MCMR method, which we used in our modelling, and those that we obtained in present work.

\section{Quantum mechanical calculations of absorption and circular dichroism from MCMR data.}
The set of data obtained from MCMR contains the excitation energy of each BChl and the transition dipole moments, see tables in appendix A. In our modelling we stick to the one-exciton Hamiltonian of the system in the site representation, (the $\hbar$ constant is set to 1 until the formula~(\ref{OD}))
\begin{equation}\label{ham}
\mathcal{H}=\sum_i \eps_{i}|i><i| + \sum_{i\ne j} M_{ij}|i ><j|.
\end{equation}
The $\eps_i$ is the energy of the $i$th BChl unit in the excited state shifted by the ground state energy, i.e. the excitation energy of the $i$th  BChl unit. Vectors $|i>$ stay for the eigenvectors of the exited state of the $i$th unit. Parameters $M_{ij}$ are the interaction energy between the $i$th and $j$th sites. Here only the terms responsible for the energy exchange are taken into account, while the static Coulomb interaction is considered as negligible. The quantities $M_{ij}$ are the Coulomb matrix elements describing the excitation transfer between the units. As in many Coulomb potential models these terms can be approximated by a few terms of multipole expansion. The main dipole-dipole contribution is denoted below by the same letter $M_{ij}$, while the dipole-quadrupole correction by the $\Delta M_{ij}$.

The antenna system linearly interacts with the external electric field oscillating in time, which in rotating wave approximation is represented as ${\bm E} e^{-\i \omega t}$. The field is spatially uniform on the molecule sizes, so that the position dependence is neglected. The corresponding dipole operator matrix elements are denoted as ${\bm d}_i$ for each unit $i$. As soon as our system is opened one has to employ the probabilistic language of the density matrix to its description. Note, that the opening of the system~(\ref{ham}) also assumes interaction with the protein degrees of freedom. To this end a number of approaches have been proposed, see for instance~\cite{NG2013, CJCCGKGCM2016,SF2000}. Here,  however, we neglect such effects in order to formulate some ''clean`` model first.  

\subsection{Formulas for OD and CD}
Evolution of the density matrix satisfies the equation $\dot \rho=-\i[\mathcal H,\rho]-\frac{1}{2}(\Gamma\rho+\rho\Gamma)$, where $\Gamma$ is a phenomenological (within our basic description) matrix of decay rates. It is responsible for broadening of the spectral peaks. The evolution equation, due to the commutator of the Hamiltonian~(\ref{ham}) with the density matrix, includes only the excitation energies of each pigment, $\eps_i$. Therefore the elements of the density matrix define the probability of the $i$th unite to be excited ($\rho_{ii}$), or be in a coherent superposition with the excited state on the $j$th unit ($\rho_{ij}$) or with the common ground state ($\rho_{i0}$). Note, that in the method of hierarchical density matrix~\cite{NG2013,IF2009,IF2009a} the total molecular Hamiltonian includes the bath degrees of freedom, so that the density matrix equation can be written as a hierarchical set of equations for zero, one, two and further numbers of excited phonon modes. In addition to the excitation of the electron density of the pigments one should take into account the reorganisation energies of the molecular skeleton in the potential well formed by the electronic density, below this term is denoted as $\lambda$. Here, within our basic consideration, we consider situation of zero phonons and omit, for a while, the reorganisation energy terms. Following the standard recipe, we separate the fast and slow degrees of freedom by introducing $\rho_{i0}=\sigma_{i0}e^{-\i \omega t}$ for all density matrix elements describing superposition between the $i$th excited state and the ground state, and consider the linear response on the weak monochromatic electric field. The steady state solution satisfies the system of equations (prime at the summation symbols means excluding the terms with the identical indices)
\begin{eqnarray*}
\left(\eps_{i}-\eps_{j} -\i\gamma_{ij}\right)\rho_{ij} + {\sum_{k>0}}^\prime (M_{ik}\rho_{kj}-\rho_{ik}M_{kj})&=&{\bm d}_i{\bm E}\sigma_{0j} - \sigma_{i0}{\bm d}_j{\bm E},\quad i\ne j\\
-\i\gamma_{ii}\rho_{ii} + {\sum_{k>0}}^\prime (M_{ik}\rho_{ki}-\rho_{ik}M_{ki})&=&{\bm d}_i{\bm E}\sigma_{0i} - \sigma_{i0}{\bm d}_i{\bm E},\quad i>0\\
\left(\eps_{i}-\omega-\i\gamma_{i}\right)\sigma_{i0} + {\sum_{k>0}}^\prime M_{ik}\sigma_{k0}&=&
(\rho_{00}-\rho_{ii}){\bm d}_i{\bm E} - {\sum_{k>0}}^\prime\rho_{ik}{\bm d}_k{\bm E},\quad i>0\\
-\i\sum_{i>0}\gamma_{ii}\rho_{ii}&=&\sum_{k>0}(\sigma_{0k}{\bm d}_k{\bm E}-\sigma_{k0}{\bm d}_k{\bm E}),
\end{eqnarray*}
Assuming the initial state to be $\rho_{ij}(t=0)=\delta_{i0}\delta_{j0}$ in the zeroth order of perturbation the solution is $\rho_{ij}(t=0)=\delta_{i0}\delta_{j0}$. Thus the first order equations for $\sigma_{i0}$ is
\begin{equation}
\left(\eps_{i}-\omega-\i\gamma_{i}\right)\sigma_{i0}+{\sum_k}^\prime  M_{ik}\sigma_{k0}
={\bm d}_i{\bm E}\label{sigma}\\
\end{equation}  
The susceptibility tensor can be read off from the expression for polarization $\bmP=\hbar\tr (\rho {\bm d})$, it is ($a,\,b=x,y,z$)
\begin{equation}\label{chi}
\chi_{a,b}(\omega)= \sum_{i,j}{\bm d}_{i}^a G_{ij}(\omega){\bm d}^b_j,
\end{equation}
where the response function $G_{ij}(\omega)$ is the inverse matrix solving the matrix  equation~(\ref{sigma}). The absorption spectrum (OD) can be obtained by averaging of the susceptibility imaginary part over all orientations of the molecular system (here we again write $\hbar$ explicitly),
\begin{equation}\label{OD}
\mathrm{OD}(\omega)=\frac{4\pi\omega}{3\hbar c}\Im{\sum_{i,j}({\bm d}_{i}\cdot{\bm d}_j) G_{ij}(\omega) }.
\end{equation}  
The averaged over orientations circular dichroism (CD) spectrum can be calculated by the formula
\begin{equation}\label{CD}
\mathrm{CD}(\omega)=\frac{4\pi\omega k_\omega}{3\hbar c}\Im{\sum_{i,j}\big({\bm d}_{i}\times {\bm d}_j\cdot (\bm R_j-\bm R_i)\big) G_{ij}(\omega) }.
\end{equation} 
In the above formulas $\cdot$ denotes scalar and $\times$ vector products, $k_\omega$ is the wavevector and $\bm R_j$ is the radius-vector of the $j$th transition dipole moment center, i.e. the Mn atom in the $j$th BChl unit.

\subsection{The interaction constants $M_{ij}+\Delta M_{ij}$}
In our consideration the formulas~(\ref{OD}),~(\ref{CD}) include only the parameters of the Hamiltonian, that can be directly calculated by the method MCMR and a set of phenomenological parameters incorporated into the decay rates matrix $\Gamma$. In fact, it is natural to assume that the decay into the ground state is the same for all sites and in our calculations we set all $\gamma_{i}$ to be 53cm$^{-1}$, as reported in~\cite{SF2000}. For the parameters, at hand, the elements of the matrix $M$ are calculated in the dipole approximation,
\begin{equation}\label{dipoleInt}
M_{ij}=C d_i d_j\frac{({\bf n}_i\cdot{\bf n}_j)\abs{{\bf r}_{ij}}^2-3({\bf n}_i\cdot{\bf r}_{ij})({\bf n}_j\cdot{\bf r}_{ij})}{\abs{{\bf r}_{ij}}^5}.
\end{equation}
The vectors $\bf n$ are the unit vectors pointing in the direction of the transition dipole moments, $\bm d_i=d_i\bm n_i $ and ${\bf r}_{ij}$ is a vector connecting the centres of $i$th and $j$th units. If the distances ${\bf r}_{ij}$ are taken in \AA, and the transition dipole moments in e\AA, then the value of $C$, which is a combination of fundamental constants, should be equal to $72973.5$ to give the values of $M_{ij}$ in~cm$^{-1}$. 

As we have noticed in the introductory part, it is important to account the dipole-quadrupole terms of the multipole expansion, their explicit formula is
\begin{multline}\label{dipoleQInt}
\Delta M_{ij}=C d_i\frac{\frac{5}{2}(\bm n_i\cdot \bm r_{ij} )(\bm r_{ij}\cdot\hat Q_j \bm r_{ij})- \abs{\bm r_{ij}}^2 (\bm n_i\cdot \hat Q_j \bm r_{ij})}{\abs{\bm r_{ij}}^7}\\-
C d_j\frac{\frac{5}{2}(\bm n_j\cdot \bm r_{ij} )(\bm r_{ij}\cdot \hat Q_i \bm r_{ij})- \abs{\bm r_{ij}}^2 (\bm n_j\cdot \hat Q_i \bm r_{ij})}{\abs{\bm r_{ij}}^7}.
\end{multline}
The term~(\ref{dipoleQInt}) includes the transition quadrupole moments tensors $\hat Q_i$. Note, that since the dipole moment is not zero, the quadrupole moment tensor depends on the choice of the coordinate system. Their values, measured in the units e\AA$^2$, were calculated from the MCMR approach, see the appendix B. 

Generically, calculation of the spectral characteristics for an isolated dimer is based on less or more accurate calculation of the Coulomb interaction integral (see the review of methods in~\cite{KK2016}).

\subsection{Computational details for the MCMR method}

Excitation energies and transition dipole moments for all
bacteriochlorophylls contained in LH2 were
obtained as previously described,\cite{Luca} whereas transition
quadrupole moments were explicitly computed for this work.
Briefly, we performed single point energy calculations at the
SA-RASSCF/MS-RASPT2 level of theory.
This implies to compute 2 roots, state average (SA), restricted
active space self-consistent field (RASSCF) energies, followed by a second order perturbation
theory correction (the PT2  part).

RASSCF is a multiconfigurational method that represents an extension of
CASSCF (complete active space self-consistent field).\cite{casscf}
The multiconfigurational character is achieved by a linear combination
among all possible configurations obtained by distributing active
electrons in a set of active orbitals.
In the specific case, we employed an active space comprised of 11 RAS1
orbitals (fully occupied in all configurations, but for a certain
amount of possible holes), 4 RAS2 orbitals (having any possible
occupation), 10 RAS3 orbitals (always empty, but for a number of
allowed excited electrons); these orbitals were occupied by 26 electrons,
and we allowed 3 holes in RAS1 and 3 excitations in RAS3.
This active space span the entire $\pi$ system of bacteriochlorophyll.

The RASSCF wavefunction was employed as reference for the following PT2
correction, using the MS-RASPT2 method.
RASPT2\cite{raspt2} and CASPT2\cite{caspt2_1990, caspt2_1992} are the
standard methods to apply a second-order M{\o}ller-Plesset perturbation to a RASSCF and
CASSCF reference wavefunction, respectively.
The multistate treatment performs a further diagonalization of the
obtained energies.\cite{ms_caspt2, Finley1998}

All calculations employed a double-zeta ANO-RCC basis set,\cite{ANO-RCC}
and Cholesky decomposition.\cite{cholesky, cholesky-casscf, cholesky-caspt2}
For MS-RASPT2 calculations we deleted the 300 highest in energy virtual
orbitals, to save on computation time, employed default IPEA
shift,\cite{ipea} and an
imaginary level shift value of 0.1.\cite{Forsberg1997}

All calculations were performed using MOLCAS [ver. 7.8].\cite{molcas7}
The calculation of the transition quadrupole moments is a newer feature,
which is present in MOLCAS [ver. 8.2].\cite{molcas8}

\section{Results} To demonstrate the properties of the model we start our consideration from a ''clean`` case, where no other parameters except those obtained from the MCMR are used. On the next step of sophistication we introduce the reorganisation energy $\lambda$ and formulate the minimal model for calculation of OD and CD spectra, which also includes the dipole-quadrupole interaction terms. To check the influence of the above dipole-quadrupole contribution, we have introduced a simplified model of charge distribution within the BChl unit, such that the transition quadrupole moments can be parametrised.

\subsection{OD and CD spectra in the ''clean`` case and for the minimal model.}
The excitation energies calculated from MCMR are in good agreement with those reported in literature. Namely, in~\cite{CJCCGKGCM2016} the excitation energies calculated in vacua, i.e. without the influence of polarizable environment, are 13634, 13631, 13633 cm$^{-1}$ for $\alpha$, $\beta$ and $\gamma$ units, respectively (compare with the values 13550, 13369, 13631 from the appendix A).    

\begin{figure}
\includegraphics[scale=0.4]{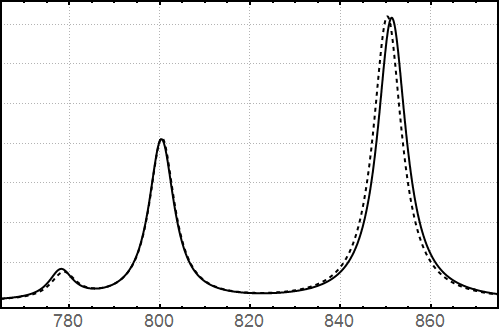}\hspace{10pt}
\includegraphics[scale=0.4]{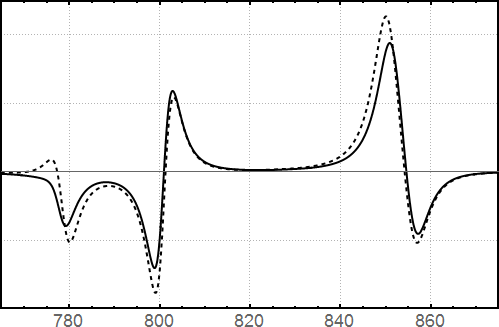}

\caption{\Small Spectra calculated for the minimal model. {\it Left panel:} The absorption spectra (arbitrary units vs the wave-length in nm) with (solid line) and without (dashed line) the dipole-quadrupole contribution to the interaction energy. {\it Right panel:} The circular dichroism spectra with (solid line) and without (dashed line) the dipole-quadrupole contribution. }\label{Fig1}
\end{figure}

As expected the  OD and CD spectra obtained for the ''clean`` case (i.e. only those that are based on the calculated from MCMR data) are blue shifted, but contain two well separated regions. The blue shift of the minor absorption peak from the experimental value is around 65 nm. To match the experimentally observed peak in the absorption spectrum we have introduced a reorganization energy $\lambda$ into the model. The suitable value for $\lambda$ was found to be 1110 cm$^{-1}$.  The corresponding spectra are shown on the fig.~\ref{Fig1}. The highest peak position is sensitive to the presence of quadrupole moment, since it corresponds to the close-packing $\alpha-\beta$ ring. The positions are 851.35 nm with $\Delta M_{ij}$ and 850.41 without it. The CD spectra cross the zero line at points 800.96 and 854.62 nm with $\Delta M_{ij}$ and at 801.28 and 854.18 nm without the dipole-quadrupole correction. The maximal deviation of the CD spectra amplitude is around~20\%.

\subsection{The energies of excitonic states and total transition dipole moments}
\begin{figure}
\includegraphics[scale=0.5]{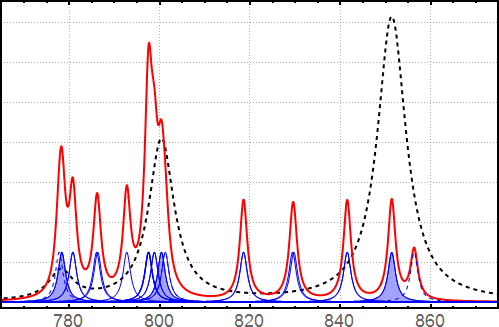}\vspace{10pt}\\
\includegraphics[scale=0.5]{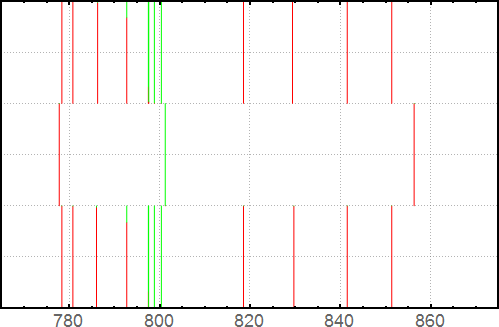}
\caption{\Small {\it Top panel:} Density of states (solid red line), calculated by~(\ref{DOS}) with $\gamma=1$ cm$^{-1}$. The absorption spectrum line (dashed) is given for comarison. The shapes drawn with thin blue line represent distribution of 27 individual exciton states, one can distinguish only 13 of them. {\it Bottom panel:} Diagram of all excitonic states. The color depicts the value of the probability~(\ref{rprob}), the length of the red part is proportional to $r_k$, while the green part length corresponds to $1-r_k$. The maximally mixed states have the value of $r_k$ (or $1-r_k$) around~0.16.}\label{Fig2}
\end{figure}

 On the figure~\ref{Fig2} we demonstrate the positions of the energy levels and the density of states of the excitonic Hamiltonian~(\ref{ham}) with the $\Delta M_{ij}$ term included. The states are spread over a significant spectral span, while there are three pairs of optically active states only. To distinguish them we have calculated the total transition dipole moments for each state. Matrix $G_{ij}(\omega)$ appearing in the formulas~(\ref{chi}),~(\ref{OD}) and~(\ref{CD}) can be diagonalised by orthogonal matrix $O$ to give the expression:
 $$
  G_{ij}(\omega)=\sum_{k} O_{ki} \frac{1}{\tilde{\eps}_k-\omega-\i\gamma_k}O_{kj},
 $$
where $\tilde{\eps}_k$ is the $k$th energy level of the excitonic Hamiltonian. One usually index these levels by $0,\pm 1,\dots \pm 7, 8$ for $\alpha-\beta$ ring and the similar set $0,\pm 1,\dots,\pm 8$ for the $\gamma$ ring. The the vector $\bmD_k$ of the total dipole moment corresponding to the $k$th level can be calculated as the vector sum $\bmD_k=\sum_j O_{kj}\bm d_j$. In addition we characterise the $k$th wave-function by the quantum probability $r_k$ to find the exciton on the $\alpha-\beta$ ring, i.e. the fraction of the wave-function localized on the B850 ring,
\begin{equation}\label{rprob}
r_k=\sum_{j\in \{\alpha-\beta\;\mbox{\scriptsize  ring}\}}O_{kj}^2=1-\sum_{j\in \{\gamma\;\mbox{\scriptsize  ring}\}}O_{kj}^2.
\end{equation}
The details of the spectrum, the total dipole moments and the calculated participation probabilities are gathered in the appendix B. The density of states (DOS) function (see figure~\ref{Fig2}) is the one calculated by the formula
\begin{equation}\label{DOS}
DOS(\omega)=\Im{\sum_{k}\frac{1}{\tilde{\eps}_k-\omega-\i\gamma_k}}.
\end{equation}
The poles of the DOS function plays a major role at consideration of the quantum transport processes in the molecular complex. Here, mostly for presentation purposes, we replaced the value of $\gamma_k$, which was 53 cm$^{-1}$, by 1 cm$^{-1}$.    

\begin{figure}
\includegraphics[scale=0.3]{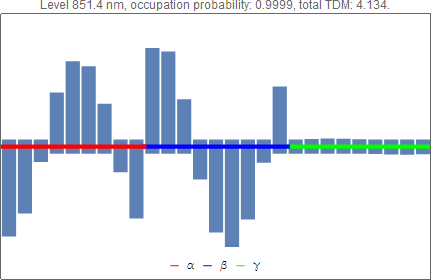}\hspace{10pt}
\includegraphics[scale=0.3]{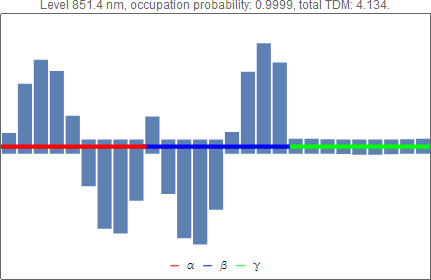}\vspace{10pt}\\
\includegraphics[scale=0.3]{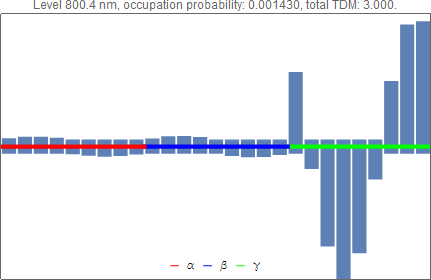}\hspace{10pt}
\includegraphics[scale=0.3]{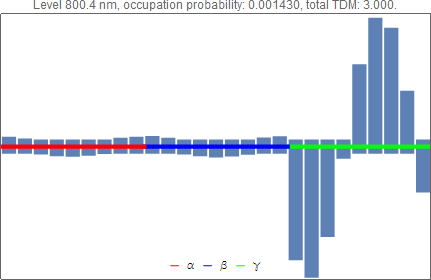}\vspace{10pt}\\
\includegraphics[scale=0.3]{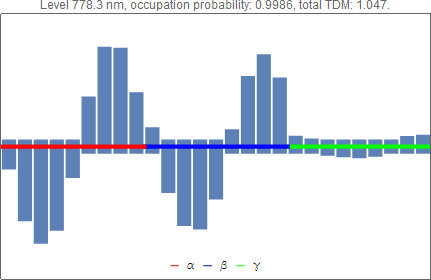}\hspace{10pt}
\includegraphics[scale=0.3]{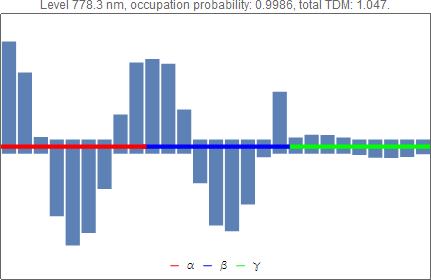}
\caption{\Small {\it Top panels:} Two wave-functions of the optically active states localised at B850 ring corresponding to the energy level 850 nm. {\it Second from the top panels:} Two wave-functions of the optically active states localised at B800 ring corresponding to the energy level 800 nm. {\it Bottom panels:} Two wave-functions of the optically active states localised at B850 ring corresponding to the energy level 780 nm.}\label{Fig3}
\end{figure}

As one can see from the figure~\ref{Fig2}, the optically active states are localized on one or the other ring. There are two states having the energy around the 850 cm$^{-1}$ with the largest total transition dipole moment. The corresponding wave-functions occupies mainly the B850 ring and both have odd symmetry, see figure~\ref{Fig3}. One is rotated by $\pi/4$ with respect to the other one. As one can see from the figure the $\alpha$ and $\beta$ units transition dipole moments come with opposite signs, which increases the total transition dipole moment. The similar picture can bee seen for the the optically active states localized on the $\gamma$ ring, see information on the figure~\ref{Fig3}. To compare the obtained wave-functions with the other possibilities we present two more pictures of the wave-function generating the minor absorption peak at 780 nm (figure~\ref{Fig3}) and of those that correspond to the quasi-mixed states, see figure~\ref{Fig6} and explanations in the next section.

The averaged absorption spectrum contains only diagonal elements of the susceptibility tensor~(\ref{chi}). The function~(\ref{CD}) contains more detailed information about the states, since it is sensitive to the particularities of the tensor $\chi_{a,b}$. The main contribution to this tensor comes from those total transition dipole moments that are not collinear. Such situation happens for all pairs of states with odd symmetry, which total transition dipole moments lay in the $x-y$ plain. The total transition dipole moments of even symmetry states are aligned in the $z$ direction, see table in appendix B for more details. In appendix B we also present the plots of different components of the susceptibility tensor $\chi_{ab}$. Among all off-diagonal elements the $\chi_{xy}$ has the most pronounced maximum, which is situated at around 850 nm. The reason is the maximal value of the total transition dipole moments $\bmD_{\pm 1}$ of the corresponding wave-functions and their parity, which eventually leads to the maximal value of the vector product $\bmD_{+ 1}\times \bmD_{-1}$ entering the equation~(\ref{CD}).

\subsection{Influence of the dipole-quadrupole interaction on the structure of excitonic wave-functions.}
\begin{figure}
\includegraphics[scale=0.4]{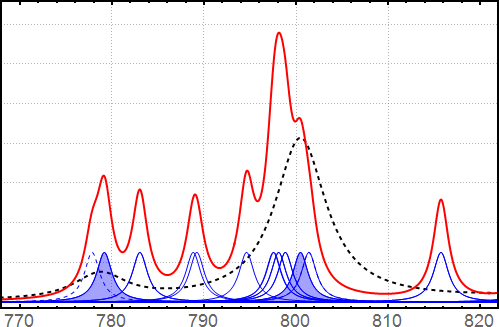}\hspace{10pt}
\includegraphics[scale=0.4]{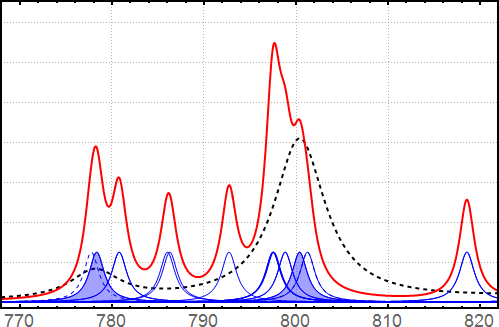}
\caption{\Small Density of states (solid red line), calculated by~(\ref{DOS}) with $\gamma=1$ cm$^{-1}$ without (left) and with (right) the dipole-quadrupole term. The absorption spectrum line (dashed) is given for comarison. The shapes drawn with the thin blue lines represent distribution of individual exciton states. }\label{Fig32}
\end{figure}

\begin{figure}
\includegraphics[scale=0.5]{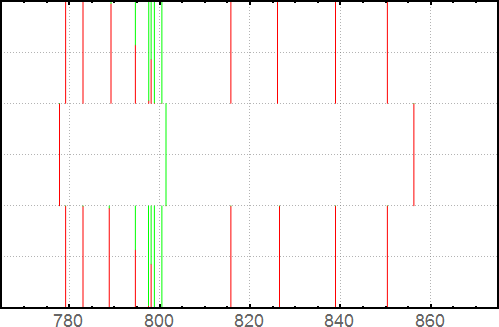}
\caption{\Small {\it Middle panel:} Diagram of all excitonic states in case of {\it zero dipole-quadrupole} term. The color depicts the value of the probability~(\ref{rprob}), the length of the red part is proportional to $r_k$, while the blue part length corresponds to $1-r_k$. The diagram demonstrates presence of the states delocalized over both rings around 800 nm, compare with the bottom plot on the figure~\ref{Fig2}.}\label{Fig22}
\end{figure}

\begin{figure}  
\includegraphics[scale=0.3]{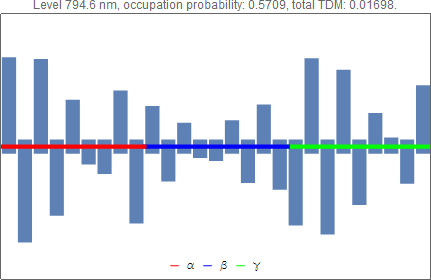}\hspace{10pt}
\includegraphics[scale=0.3]{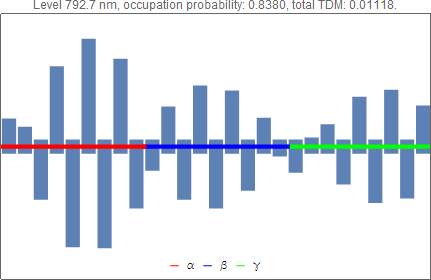}\includegraphics[scale=0.3]{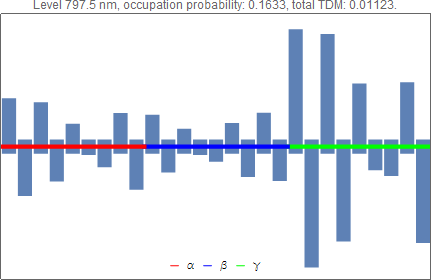}
\caption{\Small {\it Left panel:} The wave-function of the state delocalised over both B800 and B850 rings corresponding to the energy level 794 nm in case {\it without the dipole-quadrupole} term. {\it Two right panels:} Two wave-functions with the energy levels close to 794 nm calculated in case {\it with the dipole-quadrupole} term. There are no essentially delocalised states when the dipole-quadrupole term is on.}\label{Fig42}
\end{figure}

An interesting effect caused by the dipole-quadrupole interaction cannot be detected from the linear spectra, but might be significant for the non-linear responses. The difference between two cases, on and off $\Delta M_{i,j}$ term, can be seen already from the DOS plots on the figure~\ref{Fig32}. While the optically active levels stays almost intact to the additional term, some of the dark states are shifted from their places on the values of around 4nm. Moreover, the dipole-quadrupole term changes the structure of the wave-functions. Namely, the occupation probabilities depicted on the figure~\ref{Fig22} shows that there are four excitonic levels with almost equal occupation of both rings when the term is off, while the rings becomes almost separated when it is on, see figure~\ref{Fig2}. The structure of one of the delocalized wave-functions for the off case and two closest wave-functions for the on case is shown on the figure~\ref{Fig42}. Such behaviour of the wave-functions can be explained on heuristic level by the following argument. The distances between the nearest $\alpha$ units and the nearest $\alpha$ and $\gamma$ units are almost equal, about 17 \AA. The same is true for the nearest $\beta$ and $\beta$-$\gamma$ units. This quasi-symmetry in the excitonic Hamiltonian results in a special set of states with the wavelength of order 20 \AA oscillating on both rings. Introduction of the quadrupole term destroys this symmetry and lead to separation of the rings, such that the maximal values of $r_k$ and $1-r-k$ becomes not larger then~17\%.


\subsection{Static disorder in orientation of the quadrupole moments. An electrostatic model of BChl pigment.}
To study influence of the static disorder in distribution of the the quadrupole moments we provide a simplified electrostatic model of the BChl pigment. In the model the real electronic density is replaced by six charges placed in the vertices of a unit octahedron. The diagonals of the octahedron oriented over the principal axes of the transition quadrupole moment $\hat Q_i$ calculated from MCMR for the $i$th unit. The charges are subject to be determined from the set of the following linear equations: If three principal values $\lambda_a$ of the quadrupole moment are arranged in the order $\lambda_1>\lambda_2>\lambda_3$, then
\begin{eqnarray}\label{Qch}
2 (q_1 + q_2) - (q_3 + q_4) - (q_5 + q_6)&=&\lambda_1;\\ 
- (q_1 + q_2) + 2(q_3 + q_4) - (q_5 + q_6)&=&\lambda_2;\\ 
- (q_1 + q_2) - (q_3 + q_4) + 2 (q_5 + q_6)&=&\lambda_3; 
\end{eqnarray}
These equations supplemented by the condition of the zero total charge, i.e. 
\begin{equation}\label{ZeroCh}q_1 + q_2 + q_3 + q_4 + q_5 + q_6=0\end{equation} and the condition that the dipole of such six charges system should be equal to the corresponding transition dipole moment $\bm d_i$ give rise to a closed system of equations for the sought charges. The equation for the latter condition is
\begin{equation}\label{dip}
q_1-q_2=[\bm d_i]_1,\quad q_3-q_4=[\bm d_i]_2,\quad q_5-q_6=[\bm d_i]_3,
\end{equation}

where the symbol $[\bm d]_a$ denotes projection of the vector $\bm d$ on the $a$th principal axis.
The particular realisation of the octagon for the 301st $\alpha$ unit is shown on the fig.~\ref{Fig5}, the results of other calculations for different types of BChl units are given in appendix B. Note that the obtained model gives both the transition dipole moments and quadrupole moments, such that their fluctuations can be described by small variations of the charges values. For instance, according to the formula~(\ref{dip}) the value of the dipole moment can be changed by variation of the differences of the conjugated charges and the quadrupole moment by variation of their sums.

\begin{figure}
\includegraphics[scale=0.18]{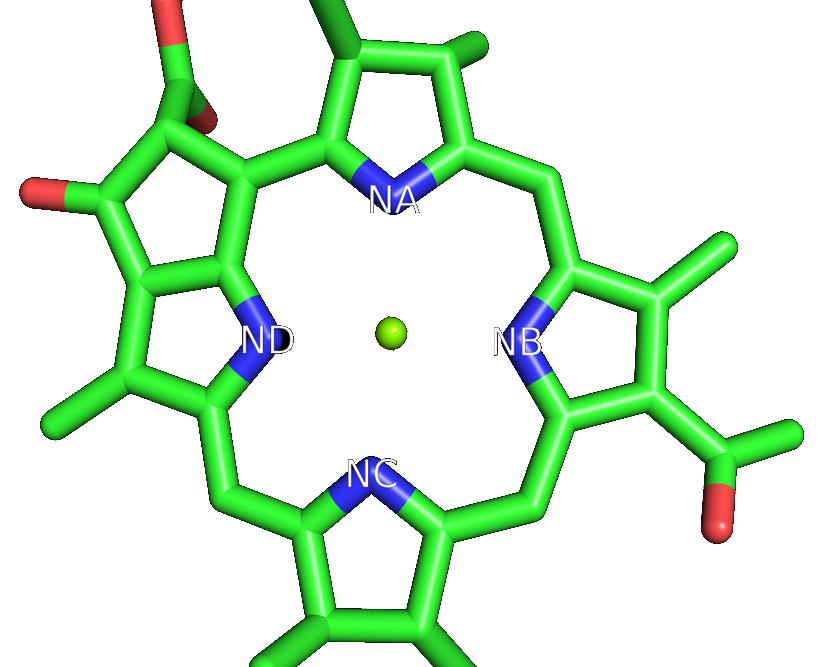}\hspace{20pt}\includegraphics[scale=0.2]{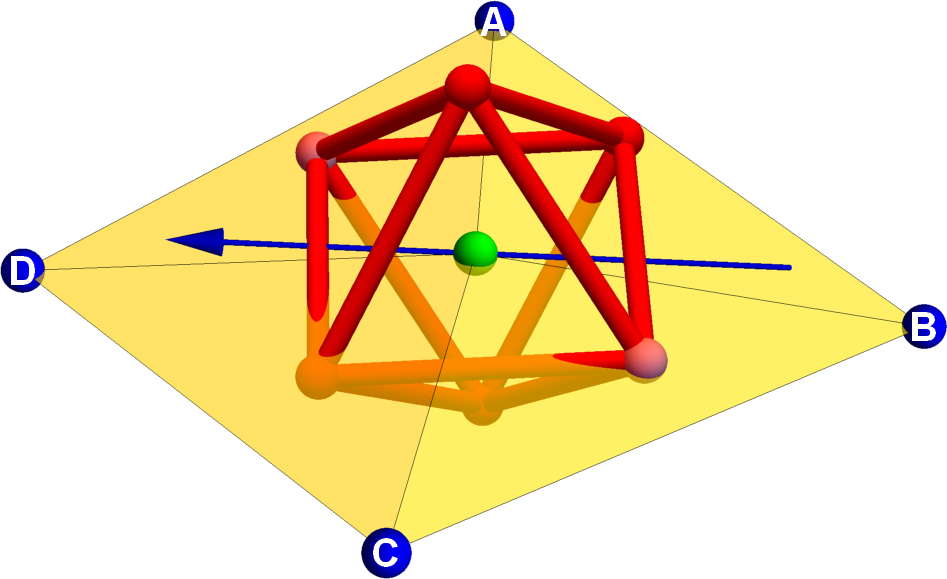}
\caption{\Small The chemical scheme of BChl unit (right) and the geometry of charges of modelled $\alpha$  unit (left). The blue spheres denote the positions of nitrogene atoms, the octagone center coinsides with the position of Mn atom (green sphere). The artificial charges are placed in the octagone vertices.}\label{Fig5}
\end{figure}

\begin{figure}
\includegraphics[scale=0.35]{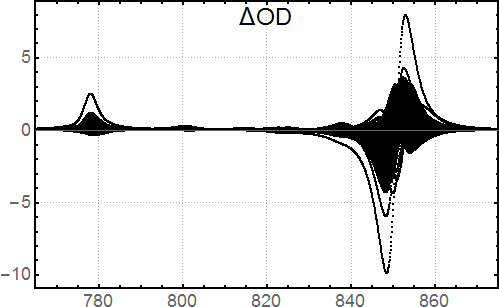}\hspace{5pt}\includegraphics[scale=0.35]{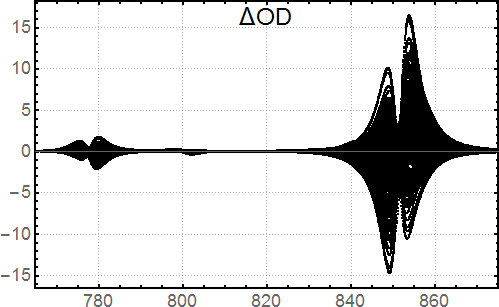}\vspace{10pt}\\\hspace{15pt}\includegraphics[scale=0.33]{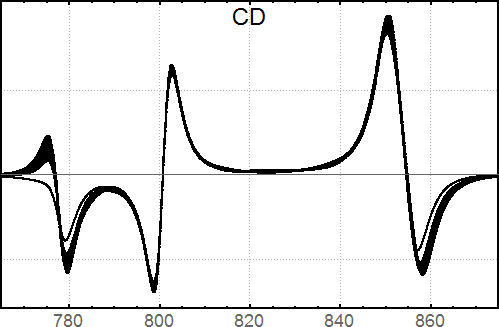}\hspace{15pt}\includegraphics[scale=0.33]{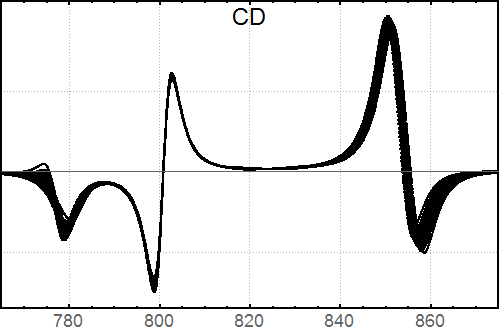}
\caption{\Small {\it Top panels:} The deviations in absorption spectrum for random rotation of the quadrupole axes around the dipole axis on the left and random deviations of the quadrupole moment strength on the right. The rotations are independent for each BChl unit with the standard deviation 7 degrees. The standard deviation of the modelled charges is 0.6 e. The values of the OD spectra deviation are given in procents of the maximal peak on both plots. {\it Bottom panels:} the CD spectra for the same ensambles.}\label{Fig6}
\end{figure}

The model of quadrupole allowed us to investigate the averaged influence of the dipole-quadrupole correction. We did two numeric experiments, first we assumed that the octahedron rotates slightly around the dipole moment direction. The rotations for each pigment are chosen to be statistically independent and distributed with respect to the Gaussian distribution with the standard deviation of 7 degrees. The plot of possible deviations of the absorption spectra from the perfectly ordered one is given on the figure~\ref{Fig6}. Both curves show that the deviations in spectra take place in the regions corresponding to the states localized on the B850 ring. As one can see the disorder in orientation of the quadrupole moments destroys the symmetry of the Hamiltonian, such that the CD spectrum returns back to the situation similar to the one when the dipole-quadrupole term is off, compare with figure~\ref{Fig1}. In the second experiment the strength of the quadrupole moment varies randomly for each unit, even strong deviations of the modelled charges cannot turn situation to the case without dipole-quadrupole interaction.

\section{Discussion}
In the article we formulated a minimal model of the LH2 complex, which reproduces the absorption and circular dichroism spectra. The model is based on {\it ab initio} calculations of the excitonic Hamiltonian and includes all interactions between the BChl units taken up to the dipole-quadrupole term. The value of the one free parameter $\lambda$ is dictated by position of the main peaks in the absorption spectrum. We interpreted $\lambda$ as reorganisation energy. Its value is 3 -- 4 times higher then the one deduced from experiment~\cite{RCRF2011}. The explanation of such discrepancy is that the Hamiltonian parameters were calculated in vacuum conditions and do not take into account polarization of the surroundings.

 The influence of this extra dipole-quadrupole term, which usually do not counting in such models, is considered in details. Main observation, we made, is that this term breaks a symmetry between the two BChl rings and makes the system effectively separated onto two subsystems. The states taking part in this phenomena are dark states and cannot influence the OD and CD spectra, while should be taken into account at consideration of the dynamical behaviour of the system, such as modelling of 2D spectroscopy experiments. 

To study influence of disorder on the level of dipole-quadrupole interaction we formulated an electrostatic model of BChl units. It is based on the possibility to reproduce both transition dipole and quadrupole moments by only six artificial charges. This allows to introduce disorder of a special type by varying the positions and values of this charges. We observed, that disorder in the strength of transition quadrupole moments do not influence on spectral characteristics, while even tiny disorientation of the quadrupoles reproduces the situation of turned off dipole-quadrupole term. 

In general, one can claim that the minimal model formulated above is able to reproduce particularities of the absorption and circular dichroism spectra and is useful for their detailed study. It is based on purely quantum-mechanical consideration and contains only one fitting parameter. Nevertheless, it has a predictive power and thus open a door for investigation of complex antenna systems. On the other hand it does not touch thermodynamic part of the problem, such as vibrational degrees of freedom and influence of polarization of the media.

\section*{Acknowledgement} Authors thank

\appendix
\section{Tables of the LH2 Hamiltonian data}
The coordinate system for the description of the LH2 geometry is chosen such that the $z$ axis is perpendicular to the planes BChl units rings, see the figure~\ref{Fig3}. In the table: the BChl units are labelled by their crystallography indexes in column I; the position-vectors $(x,y,z)$ of Mn measured in \AA~are in the columns II; the column III contains the orientation of the transition dipole moment direction vectors in  \AA; and the IV column contains its value in e\AA; $\omega$ are the corresponding excitation energies. \vspace{10pt}

{\Small \begin{tabular}{l|ccc|ccc|c}
\multicolumn{8}{l}{\bf $\alpha$-units, $\omega=13\,550\,cm^{-1}$}  \\ \hline \hline
\multicolumn{1}{c|}{I}  & \multicolumn{3}{c|}{II} &  \multicolumn{3}{c|}{III}  & IV \\\hline

 301 & 25.289 & -5.157 & 18.096 & 0.49820 & 0.85427 & 0.14842 & 1.4336 \\

 303 & 22.678 & 12.253 & 18.109 & -0.16977 & 0.97088 & 0.16905 & 1.4368 \\

 305 & 9.485 & 23.988 & 18.100 & -0.75689 & 0.63352 & 0.16056 & 1.4254 

\end{tabular}}\vspace{10pt}

\begin{figure}[b]
\includegraphics[scale=0.4]{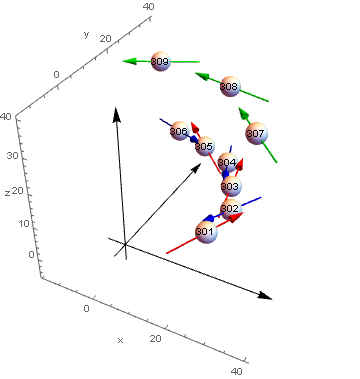}
\caption{\Small The geometry of BChl units sceleton in LH2 complex (only nine units are presented). Positions of units are shown with respect to the chosen coordinate system (the black axes). The colored arrows show the direction of the transition dipole moments.}\label{Fig4}
\end{figure}{\Small \begin{tabular}{l|ccc|ccc|c}
\multicolumn{8}{l}{\bf $\beta$-units, $\omega=13\,369\,cm^{-1}$}  \\ \hline \hline
\multicolumn{1}{c|}{I}  & \multicolumn{3}{c|}{II} &  \multicolumn{3}{c|}{III}  & IV \\\hline

 302 & 26.618 & 4.179 & 18.223 & -0.31234 & -0.93517 & 0.16706 & 1.45586 \\

 304 & 17.707 & 20.304 & 18.199 & 0.35250 & -0.91965 & 0.17317 & 1.45032 \\

 306 & 0.534 & 26.991 & 18.211 & 0.86811 & -0.47001 & 0.15960 & 1.45576 

\end{tabular}}\vspace{10pt}

{\Small \begin{tabular}{l|ccc|ccc|c}
\multicolumn{8}{l}{\bf $\gamma$-units, $\omega=13\,631\,cm^{-1}$}  \\ \hline \hline
\multicolumn{1}{c|}{I}  & \multicolumn{3}{c|}{II} &  \multicolumn{3}{c|}{III}  & IV \\\hline

 307 & 28.628 & 11.769 & 34.999 & -0.76178 & 0.63664 & 0.11992 & 1.42919 \\

 308 & 14.396 & 27.476 & 34.907 & -0.99215 & -0.00159 & 0.12505 & 1.42605 \\

 309 & -6.618 & 30.349 & 34.930 & -0.75483 & -0.64453 & 0.12172 & 1.40843 
 
\end{tabular}}\vspace{10pt}

The components of the transition quadrupole moments tensor are presented in the table below in the units e\AA$^2$. The tensors calculated in the frames positioned at the Mn atom of each BChl unit obtained by parallel translation of the original frame.  \vspace{10pt}

{\Small \begin{tabular}{l|cccccc}
\hline \hline
\multicolumn{1}{c|}{I}  & xx & yy & zz & xy &xz & yz   \\\hline

 301 & 1.7579 & -0.6932 & -1.0648 & 0.1384 & -3.272 & -5.641 \\

 302 & 1.5564 & -0.1870 & -1.3694 & 0.7644 & 2.033 & 6.233 \\

 303 & 0.7191 & 0.3412 & -1.0603 & 1.284 & 1.111 & -6.405 \\

 304 & 0.1348 & 1.1929 & -1.3276 & 1.019 & -2.325 & 5.900 \\

 305 & -0.6434 & 1.7732 & -1.1298 & 0.2685 & 4.948 & -4.162 \\

 306 & -0.3777 & 1.7410 & -1.3633 & -0.3955 & -5.677 & 2.997 \\

 307 & -3.9789 & 2.7103 & 1.2686 & 0.6745 & -0.8220 & 1.281 \\

 308 & -2.1623 & 0.8033 & 1.3590 & -3.178 & -1.517 & 0.4825 \\

 309 & 2.2164 & -3.5170 & 1.3006 & -1.860 & -1.422 & -0.5819
 
\end{tabular}}\vspace{10pt}

The coordinates and directions of the transition dipole/quadrupole moments for other units can be obtained by rotation of the coordinate system on 120 and 240 degrees about the $z$ axis.

\section{The energy levels of the excitonic Hamiltonian and structure of the wave-functions.}
The values of the energy levels of the excitonic Hamiltonian~(\ref{ham}) calculated with the parameters given in the appendix A, the reorganization energy $\lambda=1110$ cm$^{-1}$ and with the dipole-quadrupole term are gathered in the column I of the table below. The column II contains the values of the probabilities $r_k$, calculated by the formula~(\ref{rprob}). The value of the total transition dipole moment is given in column III and coordinates (in \AA) of its direction arranged in the order $x$, $y$, $z$ are in the last column. The red color in the column I denotes the states with even parity, others are odd states; in the column II red means the states localised mostly on the B850 ring, green corresponds to the B800 states the other are mixed states, finally the red color in column III marks the state with the largest total transition dipole moment.   \vspace{10pt}

{\Small \begin{tabular}{l|c|c|ccc}
\hline \hline
\multicolumn{1}{c|}{I}  & \multicolumn{1}{c|}{II} &  \multicolumn{1}{c|}{III}  &  \multicolumn{3}{c}{IV} \\\hline
\cellcolor[HTML]{FFCCC9} 777.8 & \cellcolor[HTML]{FFCCC9}0.9992 &\cellcolor[HTML]{FFCCC9} 1.01 & 0.0012113567 & -0.00023362332 & -0.99999924 \\\hline

 778.3 & \cellcolor[HTML]{FFCCC9}0.9986 &\cellcolor[HTML]{FFCCC9} 1.05 & 0.98133782 & -0.19228839 & 0.0011225040 \\

 778.3 & \cellcolor[HTML]{FFCCC9}0.9986 &\cellcolor[HTML]{FFCCC9} 1.05 & -0.19229310 & -0.98133754 & 0.00000000
\\\hline

 780.8 &\cellcolor[HTML]{FFCCC9} 0.9966 & 0.0549 & -0.035558754 & -0.99936057 & -0.0037462421 
\\

 780.8 & \cellcolor[HTML]{FFCCC9}0.9966 & 0.0545 & 0.99920254 & -0.039677299 & 0.0044722914 \\
\hline
\cellcolor[HTML]{FFCCC9} 786.0 &\cellcolor[HTML]{FFCCC9} 0.9871 & 0.0232 & -0.00051640509 & -0.0015859252 & \
-0.99999861 \\

\cellcolor[HTML]{FFCCC9} 786.3 &\cellcolor[HTML]{FFCCC9} 0.9864 & 0.0207 & -0.0073510389 & 0.0065750288 & -0.99995136 \
\\\hline

 792.7 &\cellcolor[HTML]{FFCCC9} 0.8380 & 0.0111 & -0.025241660 & 0.99968116 & -0.00065553547 \
\\

 792.7 &\cellcolor[HTML]{FFCCC9} 0.8380 & 0.0112 & 0.99974615 & 0.022249859 & 0.0035472698 \\
\hline
 797.5 &\cellcolor[HTML]{9AFF99} 0.1633 & 0.0112 & 0.71401751 & 0.70012759 & -0.00059010474 \\

 797.5 &\cellcolor[HTML]{9AFF99} 0.1634 & 0.0112 & 0.69825672 & -0.71584700 & -0.00079142351 \\
\hline
\cellcolor[HTML]{FFCCC9} 797.5 &\cellcolor[HTML]{9AFF99} 0.01498 & 0.00640 & 0.0014814248 & -0.00078268679 & -0.99999860 \\

\cellcolor[HTML]{FFCCC9} 797.6 &\cellcolor[HTML]{9AFF99} 0.01598 & 0.00890 & 0.0016730058 & -0.0015497305 & 0.99999740 \\
\hline
 798.8 &\cellcolor[HTML]{9AFF99} 0.004468 & 0.0122 & -0.99999999 & 0.00014813436 & \
-0.000021609980 \\

 798.8 &\cellcolor[HTML]{9AFF99} 0.004467 & 0.0121 & -0.0013239005 & 0.99999911 & \
0.00013241853 \\\hline

 800.4 &\cellcolor[HTML]{9AFF99} 0.001430 &\cellcolor[HTML]{FFCCC9} 3.00 & 0.13821256 & 0.99040259 & 0.00000000
\\

 800.4 &\cellcolor[HTML]{9AFF99} 0.001430 &\cellcolor[HTML]{FFCCC9} 3.00 & 0.99040255 & -0.13821287 & 0.00000000 \\\hline

\cellcolor[HTML]{FFCCC9} 801.2 &\cellcolor[HTML]{9AFF99} 0.001296 & 0.495 & -0.000012451692 & -0.000065106610 & -1.0000000 \\
\hline

 818.6 &\cellcolor[HTML]{FFCCC9} 0.9986 & 0.0143 & -0.63595626 & 0.77172502 & 0.00036280616 \\

 818.6 &\cellcolor[HTML]{FFCCC9} 0.9986 & 0.0144 & 0.77093371 & 0.63691406 & -0.0013015777 \\
\hline

\cellcolor[HTML]{FFCCC9} 829.4 &\cellcolor[HTML]{FFCCC9} 0.9978 & 0.0183 & -0.012571211 & -0.0039692012 & -0.99991310 \\

\cellcolor[HTML]{FFCCC9}  829.7 &\cellcolor[HTML]{FFCCC9} 0.9979 & 0.0307 & -0.0010384465 & 0.00040830963 & -0.99999938 \\\hline

 841.5 &\cellcolor[HTML]{FFCCC9} 0.9990 & 0.0696 & -0.14720114 & 0.98910658 & 0.000014678051 \\

 841.5 &\cellcolor[HTML]{FFCCC9} 0.9990 & 0.0702 & 0.98878018 & 0.14937788 & 0.00000000
\\\hline

 851.4 &\cellcolor[HTML]{FFCCC9} 0.9999 &\cellcolor[HTML]{FFCCC9} 4.13 & -0.95298350 & 0.30302219 & 0.00000000 \\

 851.4 &\cellcolor[HTML]{FFCCC9} 0.9999 &\cellcolor[HTML]{FFCCC9} 4.13 & -0.30302091 & -0.95298391 &  0.00000000
\\\hline

\cellcolor[HTML]{FFCCC9} 856.3 &\cellcolor[HTML]{FFCCC9} 0.9995 & 0.0829 & 0.0043513776 & 0.0099332475 & 0.99994120
 
\end{tabular}}\vspace{10pt}

The table below contains information of the quantum states of the excitonic Hamiltonian~(\ref{ham}) calculated with the parameters given in the appendix A, the reorganization energy $\lambda=1110$ cm$^{-1}$ and {\it the excluded dipole-quadrupole} term. The columns and colors have the same meaning as in the previous table.\vspace{10pt}

{\Small \begin{tabular}{l|c|c|ccc}
\hline \hline
\multicolumn{1}{c|}{I}  & \multicolumn{1}{c|}{II} &  \multicolumn{1}{c|}{III}  &  \multicolumn{3}{c}{IV} \\\hline
\cellcolor[HTML]{FFCCC9}{777.8} & \cellcolor[HTML]{FFCCC9}{0.9991} & \cellcolor[HTML]{FFCCC9}{1.01} & 0.00053882325 & -0.00010481911 & -0.99999985 \\\hline

 779.2 & \cellcolor[HTML]{FFCCC9}{0.9986} &\cellcolor[HTML]{FFCCC9}{0.983} & 0.99852945 & -0.054209117 & 0.00055830518 \\

 779.2 & \cellcolor[HTML]{FFCCC9}{0.9986} & \cellcolor[HTML]{FFCCC9}{0.983} & -0.054197594 & -0.99853023 & 0.000067531803 \\\hline

 783.0 & \cellcolor[HTML]{FFCCC9}{0.9958} & 0.0293 & -0.34390242 & -0.93900044 & -0.0030496843 \\

 783.0 &\cellcolor[HTML]{FFCCC9}{0.9958} & 0.0292 & 0.93697229 & -0.34937265 & 0.0046566681 \\\hline

\cellcolor[HTML]{FFCCC9}{  788.8} & \cellcolor[HTML]{FFCCC9}{0.9772} & 0.0176 & -0.00019128986 & -0.0020457714 & -0.99999789 \\

\cellcolor[HTML]{FFCCC9}{  789.2} & \cellcolor[HTML]{FFCCC9}{0.9748} & 0.0274 & -0.0046554240 & 0.0031184836 & -0.99998430 \\\hline

 794.6 & 0.5709 & 0.0170 & -0.31009629 & -0.95070508 & 0.00038185461 \\

 794.6 & 0.5708 & 0.0170 & -0.95047089 & 0.31080856 & -0.0017666498 \\\hline

\cellcolor[HTML]{FFCCC9}{  797.5} & 0.02481 & 0.00589 & -0.0042671905 & 0.00034652028 & 0.99999084 \\

\cellcolor[HTML]{FFCCC9}{ 797.6} & 0.02751 & 0.0106 & -0.0021784370 & 0.0021995584 & -0.99999521 \\\hline

 798.1 & 0.4301 & 0.00656 & -0.98866659 & -0.15012631 & -0.00068418116 \\

 798.1 & 0.4302 & 0.00655 & -0.14119443 & 0.98997443 & 0.0038423341 \\\hline

 798.8 & \cellcolor[HTML]{9AFF99}{ 0.005682} & 0.0125 & -0.89549333 & 0.44507493 & 0.000051510282 \\

 798.8 & \cellcolor[HTML]{9AFF99}{ 0.005678} & 0.0125 & -0.44413518 & -0.89595977 & -0.00017432100 \\\hline

 800.5 & \cellcolor[HTML]{9AFF99}{ 0.001489} & \cellcolor[HTML]{FFCCC9}{3.01} & 0.13681247 & 0.99059697 & 0.00000000 \\

 800.5 & \cellcolor[HTML]{9AFF99}{ 0.001489} & \cellcolor[HTML]{FFCCC9}{3.01} & 0.99059692 & -0.13681278 & 0.00000000\\\hline

\cellcolor[HTML]{FFCCC9}{ 801.4} & \cellcolor[HTML]{9AFF99}{ 0.001117} & 0.493 & -0.000011534575 & -0.000067651763 & -1.0000000 \\\hline

 815.7 & \cellcolor[HTML]{FFCCC9}{0.9985} & 0.0124 & -0.89307985 & 0.44989788 & 0.00052320906 \\

 815.7 & \cellcolor[HTML]{FFCCC9}{0.9985} & 0.0125 & 0.45091632 & 0.89256526 & -0.0013107850 \\\hline

\cellcolor[HTML]{FFCCC9}{  826.1} &\cellcolor[HTML]{FFCCC9}{ 0.9978} & 0.0145 & -0.013123871 & -0.0042765869 & -0.99990473 \\

\cellcolor[HTML]{FFCCC9}{  826.5} & \cellcolor[HTML]{FFCCC9}{0.9979} & 0.0333 & -0.0014563467 & 0.00072009688 & -0.99999868 \\\hline

 838.9 & \cellcolor[HTML]{FFCCC9}{0.9990} & 0.0379 & 0.13295172 & 0.99112252 &0.00000000 \\

 838.9 & \cellcolor[HTML]{FFCCC9}{0.9990} & 0.0383 & 0.99193963 & -0.12671139 & -0.000049365544 \\\hline

 850.4 & \cellcolor[HTML]{FFCCC9}{0.9999} & \cellcolor[HTML]{FFCCC9}{4.14 }& -0.97661947 & 0.21497539 & 0.00000000 \\

 850.4 & \cellcolor[HTML]{FFCCC9}{0.9999} & \cellcolor[HTML]{FFCCC9}{4.14 }& 0.21497547 & 0.97661945 & 0.00000000\\\hline

\cellcolor[HTML]{FFCCC9}{ 856.3} & \cellcolor[HTML]{FFCCC9}{0.9998} & 0.0860 & 0.0041055290 & 0.0084573043 & 0.99995581
\end{tabular}}\vspace{10pt}

\begin{figure}
\includegraphics[width=6cm, height=3.5cm]{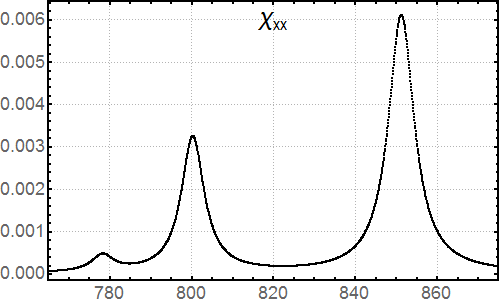}\hspace{10pt}\includegraphics[width=6cm, height=3.5cm]{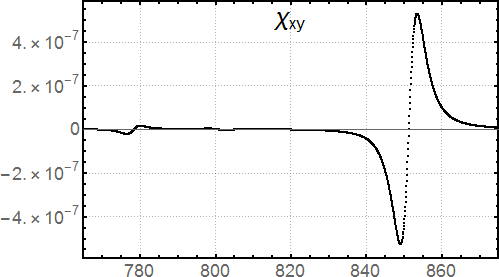}
\\
\includegraphics[width=6cm, height=3.5cm]{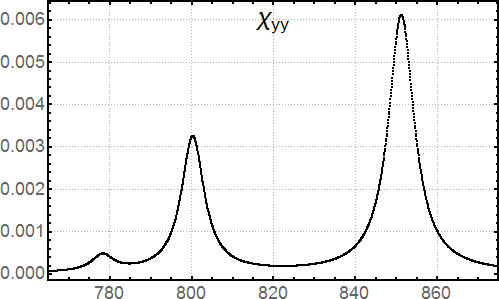}\hspace{10pt}\includegraphics[width=6cm, height=3.5cm]{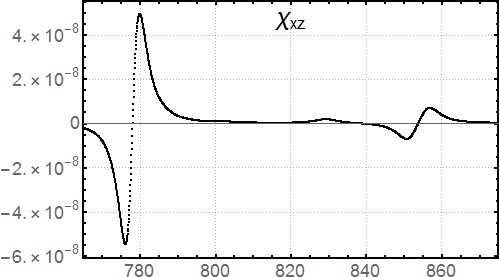}
\\
\hspace{-7pt}\includegraphics[width=6.3cm, height=3.5cm]{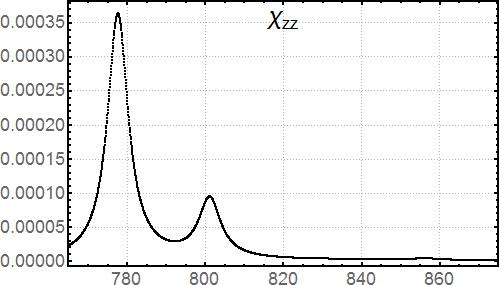}
\hspace{2pt}\includegraphics[width=6.15cm, height=3.5cm]{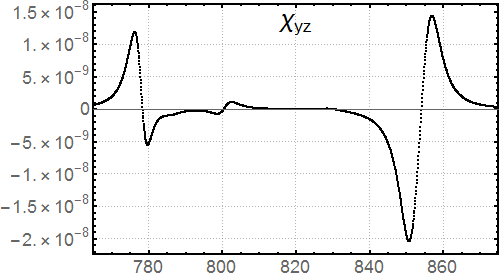}
\caption{\Small The susceptibility tensor spectral behaviour.}\label{Fig7}
\end{figure}

The components of the susceptibility tensor~(\ref{chi}) as a function of the frequency $\omega$ are presented on the plots on the figure~\ref{Fig7}.

{\it Calculation of artificial charges and their positions for the modelled transition quadrupole moment.}
The principle values of the transition quadrupole moment for the unit 301 are $\lambda_1=-7.12798$, $\lambda_2=6.17111$, $\lambda_3=0.956871$ e\AA$^2$, the transition dipole vector has the components $(0.714211,1.22467,0.212768)$ e\AA, then the values of calculated charges expressed in units of elementary charge are (see equations~(\ref{Qch})-(\ref{dip}))
$$\begin{array}{ccc}
q_1=-0.6300,& q_3=0.5789,&q_5=0.1409,\\
q_2=-1.746,&q_4=1.478,&q_6=0.1781.
\end{array}
$$
The positions of the charges in the frame of the principal axis are $r_1=(1,0,0)$,  $r_2=-r_1$,  $r_3=(0,1,0)$ ,  $r_4=-r_3$ ,  $r_5=(0,0,1)$ ,  $r_6=-r_5$. In the table below the vectors are defined in the frames positioned at the Mn atom of each BChl unit obtained by parallel translation of the original frame, the radius vectors therefore are marked by apostrophe. We reproduce values for three units only, the others can be obtained by similar way ($r_2'=-r_1'$, $r'_4=-r'_3$, $r'_6=-r'_5$). \vspace{10pt}

{\Small \begin{tabular}{l|c|c|c}
\hline \hline
 Unit &$q_1$ & $q_2$ &$r_1'$\\\hline
301& -0.6300 & -1.746 &(0.2586, 0.6336, 0.7291) \\\hline
302& -0.6293 & -1.756 &(-0.1168, -0.6550, 0.7466) \\\hline
307& -1.2769 & -0.1283 &(0.9761, -0.1276, 0.1761) \\\hline\hline
&$q_3$ & $q_4$ &$r_3'$\\\hline
301& 0.5789 & 1.478 &(-0.5073, -0.5532, 0.6607) \\\hline
302& 1.5086 & 0.5868 &(-0.3834, -0.6637, -0.6422) \\\hline
307& 1.0014 & 0.1529 &(0.02368, 0.8673, 0.4972) \\\hline\hline
&$q_5$ & $q_6$ &$r_5'$ \\\hline
301& 0.1409 & 0.1781 &(0.8220, -0.5408, 0.1783)\\\hline
302& 0.1615 & 0.1284 &(0.9162, -0.3613, -0.1736) \\\hline
307& 0.0970 & 0.1538 &(-0.2162, -0.4812, 0.8496)
\end{tabular}}\vspace{10pt} 
\end{document}